\documentclass[letter]{aa}
\usepackage[varg]{txfonts}
\usepackage{graphicx}
\usepackage{natbib}
\bibpunct{(}{)}{;}{a}{}{,} % to follow the A&A style

\begin{document}

\title{Evidence for distinct modes of solar activity}

\author{I.G. Usoskin\inst{1}
\and G. Hulot\inst{2}
\and Y. Gallet\inst{2}
\and R. Roth\inst{3}
\and A. Licht\inst{2}
\and F. Joos\inst{3}
\and G. A. Kovaltsov\inst{4}
\and E. Th\'ebault\inst{2}
\and A. Khokhlov\inst{2,5}}

\institute{Sodankyl\"a Geophysical Observatory (Oulu unit) and Physics Dept., ReSoLVE Center of Excellence, University of Oulu, Finland
\and Institut de Physique du Globe de Paris, Sorbonne Paris Cit\'e, Universit\'e Paris Diderot, UMR 7154 CNRS, F-75005 Paris, France
\and Climate and Environmental Physics, Physics Institute, and Oeschger Centre for Climate Change Research, University of Bern, Switzerland
\and Ioffe Physical-Technical Institute, 194021 St. Petersburg, Russia
\and IEPT Russian Academy of Sciences, 117997 Moscow, Russia
}

\date{}

\abstract {}
%Aims
{The Sun shows strong variability in its magnetic activity,
 from Grand minima to Grand maxima, but the nature of the variability is not fully understood, mostly
 because of the insufficient length of the directly observed solar activity records and of uncertainties related to long-term reconstructions.
Here we present a new adjustment-free reconstruction of solar activity over three millennia and study its different modes.
}
%Methods
{We present a new adjustment-free, physical reconstruction of solar activity over the past three millennia, using
 the latest verified carbon cycle, $^{14}$C production, and archeomagnetic field models.
This great improvement allowed us to study different modes of solar activity at an unprecedented level of details. }
%Results
{The distribution of solar activity is clearly bi-modal, implying the existence of distinct modes of activity.
The main regular activity mode corresponds to moderate activity that varies in a relatively narrow band between sunspot numbers
 $\approx 20$ and 67.
The existence of a separate Grand minimum mode with reduced solar activity, which cannot be explained by random fluctuations
 of the regular mode, is confirmed at a high confidence level.
The possible existence of a separate Grand maximum mode is also suggested, but the statistics is too low to reach
 a confident conclusion. }
%Conclusions
{The Sun is shown to operate in distinct modes - a main general mode, a Grand minimum mode corresponding to an inactive Sun,
 and a possible Grand maximum mode corresponding to an unusually active Sun.
These results provide important constraints for both dynamo models of Sun-like stars and investigations of possible solar influence on Earth's climate.
}

\maketitle

\section{Introduction}

Solar variability has been the subject of intense studies for a long time \citep[e.g][]{hathawayLR}, but
 the physics behind it is not yet fully understood \citep{charbonenau10}.
A particularly important question is whether the so-called Grand minima and maxima reported in historical
 records correspond to special states of the solar dynamo or result from random variability \citep[e.g.,][]{nandy11,choudhuri12,carbonell94}.
Direct records of solar activity, provided by sunspot numbers since 1610 AD \citep{hoyt98}, are too short to fully address this problem.
Reconstructing its past history over longer periods is thus crucial.
Cosmogenic radionuclides, such as $^{14}$C, provide powerful proxies \citep{beer12,solanki_Nat_04,usoskin_LR_13} able to extend
 the record several millennia back in time.
However, previous reconstructions based on such proxies required an ad-hoc calibration, that lead to much
 debate \citep{muscheler_Nat_05,solanki_Nat_05}.
In the present study, we finally overcome this ambiguity and present the first fully adjustment-free physical reconstruction of solar activity,
 using the latest carbon cycle \citep{roth13}, $^{14}$C production \citep{kovaltsov12}, and archeomagnetic field \citep{licht13} models, and
 converting $^{14}$C data into a three-millennia-long sunspot number record.
We also study the statistics of the solar activity level to search for distinct modes.

\section{Method}

\begin{figure}
\centering \resizebox{\columnwidth}{!}{\includegraphics{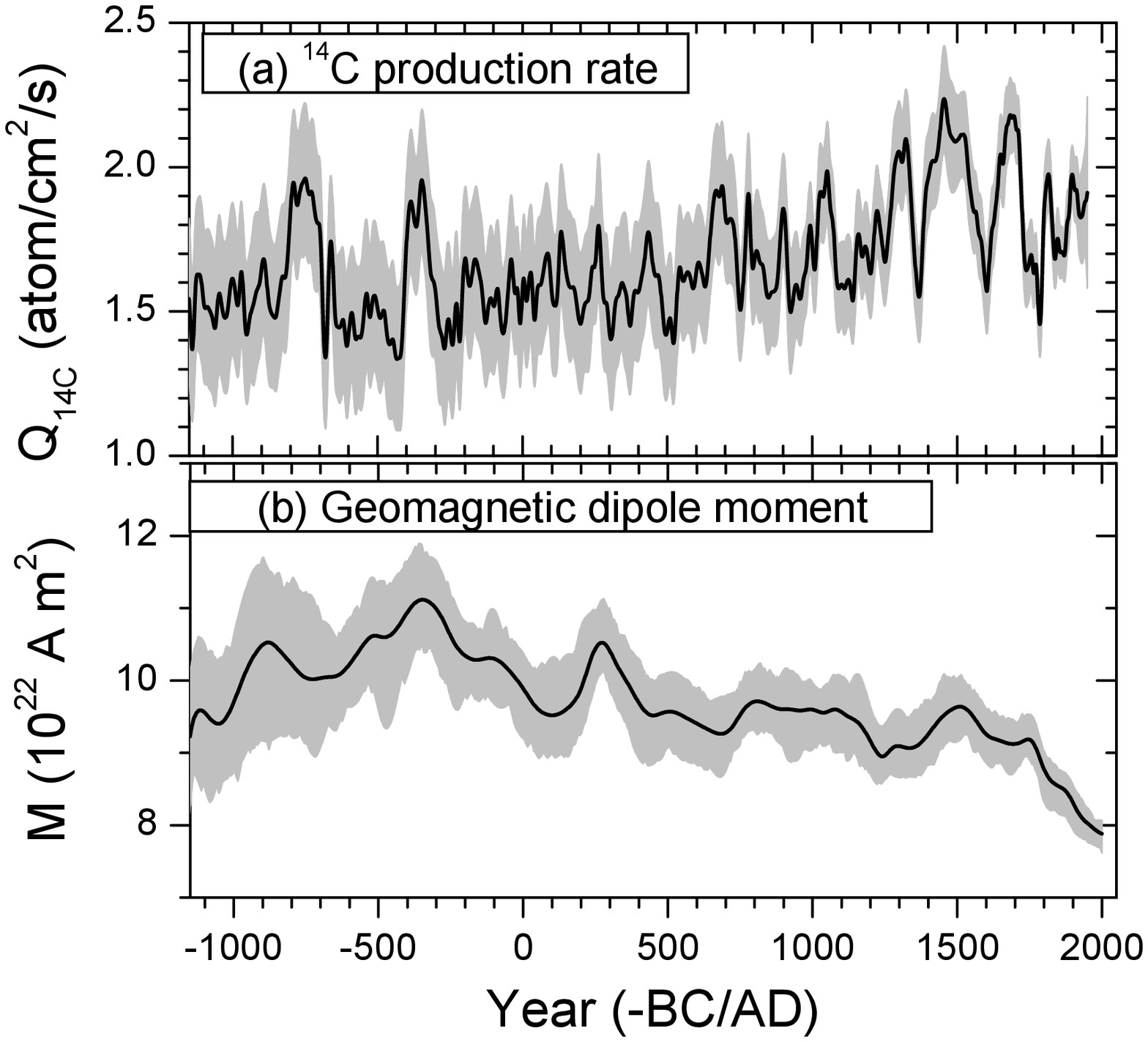}}
\caption{A) Radiocarbon $^{14}$C production rate $Q(t)$ reconstructed from IntCal09 \citep{reimer_09} and SHCal04 \citep{mccormac04} data
 in the form of an ensemble of 1000 series using
 the Bern3D-LPJ model \citep{roth13} and accounting for uncertainties;
 the mean of these series (solid line) and the inferred 95\% confidence interval (CI, shaded area) are plotted.
B) Geomagnetic dipole moment $M(t)$ as taken from the A\_FM ensemble of 1000 archeomagnetic field models \citep{licht13}, the ensemble mean (solid line) and
 inferred 95\% CI (shaded) are plotted.}
\label{Fig:data}
\end{figure}

Records of cosmogenic radionuclides stored in independently dated natural archives, such as sediments, ice cores, or tree rings,
 provide the only means to reconstruct solar activity before the beginning of direct solar observations \citep{beer12,usoskin_LR_13}.
Dendrochronologically dated $^{14}$C concentrations, in particular, prove very useful for that purpose \citep{stuiver80}, provided
 a good knowledge of the carbon cycle \citep{muscheler07}, $^{14}$C production process \citep{kovaltsov12}, and geomagnetic
 field evolution \citep{snowball07} is available.
Here we followed a standard approach \citep[e.g.][]{solanki_Nat_04}:
\begin{equation}
\Delta^{14}C(t) \stackrel{(1)}{\longrightarrow} {Q(t) + M(t)} \stackrel{(2)}{\longrightarrow} \phi(t) \stackrel{(3)}{\longrightarrow} S(t)\, .
\label{Eq:a}
\end{equation}
At each step we relied on the most recently updated models.
We used the IntCal09 \citep{reimer_09} and SHCal04 \citep{mccormac04} datasets of temporal series of well-dated
 $\Delta^{14}$C measurements over the Holocene.
Series of $\Delta^{14}$C were converted (step 1 in Eq.~\ref{Eq:a}) into a $^{14}$C production rate $Q(t)$ (Fig.~\ref{Fig:data}A, http://www.climpast.net/9/1879/2013/cp-9-1879 -2013-supplement.zip)
 using the University of Bern Earth system model of intermediate complexity Bern3D-LPJ \citep{roth13}, a new-generation carbon-cycle climate model,
 featuring a 3D dynamic ocean, reactive ocean sediments and a 2D atmosphere component coupled to
  the Lund-Potsdam-Jena dynamic global vegetation model.
$Q(t)$ was then combined with the A\_FM archeomagnetic field model $M(t)$, which is
 constructed using only archeomagnetic and volcanic paleomagnetic data \citep[][see http://geomag.ipgp.fr/download/ARCHEO\_FM.zip]{licht13},
 in the eccentric dipole approximation \citep{fraser87} (Fig.~\ref{Fig:data}B), as input (step 2) of a new model \citep{kovaltsov12} of $^{14}$C
 production by galactic cosmic rays (GCRs), parameterized in terms of a heliospheric modulation potential \citep{vainio09} $\phi (t)$.
Contribution of $\alpha-$particles and heavier species of GCRs was considered according to \citet{webber09}.
Variations in the GCR spectrum reflect solar activity since the interstellar incoming flux of GCRs can be expected to be steady on
 this time scale \citep{beer12}.
This $\phi (t)$ series was finally converted (step 3) into the sunspot number $S(t)$, using the open magnetic-flux model \citep{solanki00,krivova07}.

Uncertainties were assessed by directly applying an ensemble of 1000 time-varying individual archeomagnetic models, accounting for
 measurement and other random errors \citep{licht13}.
This ensemble was combined with a similar ensemble of 1000 production rates $Q(t)$ that accounts
 for measurement or compilation uncertainties in the IntCal09 and SHCal04 data, in the air-sea gas exchange rate,
 in the terrestrial primary production, and in the closure of the atmospheric CO$_2$ budget \citep{roth13}.
All combinations of $Q(t)$ series with archeomagnetic field models were next used to produce $10^6$ series of modulation potentials
 $\phi (t)$, then converted into $10^6$ $S(t)$ series of sunspot numbers, which reflect the error propagation.
An additional random error with $\sigma_S = 0.5$ was finally added independently to each $S(t)$ series to account
 for the small possible intermediate error from converting the modulation potential $\phi$ into a solar
 open magnetic flux $F_o$ \citep{solanki_Nat_04}.
Other systematic uncertainties are discussed in Sect.~\ref{Sec:unc}.

This allowed us to produce an ensemble of $10^6$ series of $S(t)$ that cover the period 1150 BC--1950 AD (the atmospheric $\Delta^{14}$C signal
 is affected by atmospheric nuclear bomb tests after 1950 AD) and reflect all error propagation in our reconstruction (except
 for possible systematic uncertainties, discussed below).
These series were finally averaged over ten years (calendar decades), to account for the intrinsic inability of such reconstructions to recover
 higher frequency variations.

\begin{figure}
\centering \resizebox{\columnwidth}{!}{\includegraphics{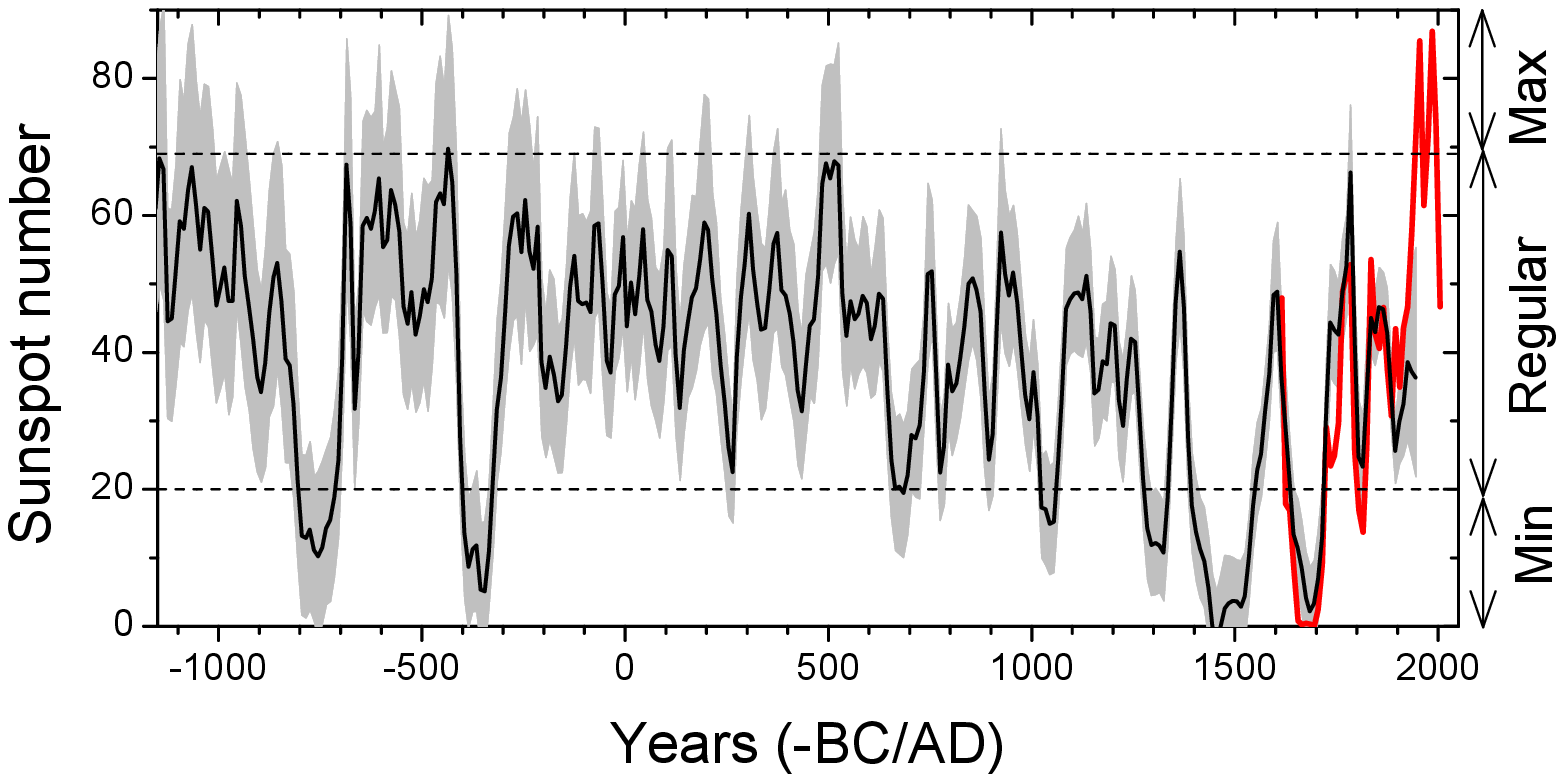}}
\caption{Series of decadal sunspot numbers, reconstructed between 1150 BC and 1950 AD, using the ensemble of $10^6$ series
 (see text); the ensemble mean (solid line) and inferred 95\% CI (shading) are shown.
Decadal group sunspot numbers \citep{hoyt98} directly observed since 1610 AD are shown in red.
Horizontal dashed lines define the bounds of the three suggested modes as defined in Fig.~\ref{Fig:dist}: Grand minimum, Regular
 and Grand maximum modes, denoted "Min", "Regular" and "Max", respectively.
 Electronic Table for this plot is available in the online material.}
\label{Fig:Res}
\end{figure}
Fig.~\ref{Fig:Res} shows the resulting mean series together with the corresponding 95\% confidence intervals (CI).
This reconstructed solar activity displays a number of distinct features, in particular well-defined Grand minima of solar activity,
 ca. 770 BC, 350 BC, 680 AD, 1050 AD, 1310 AD, 1470 AD, and 1680 AD \citep[cf. Table 1 in][]{usoskin_AA_07}.
%{\bf
Despite uncertainties in the directly observed sunspot numbers before 1848 \citep{svalgaard12,leussu13},
 remarkable agreement is found
%}
 with the decadal group sunspot numbers \citep{hoyt98} that were directly observed since 1610 AD (also shown),
 and indicates that the modern Grand maximum (which occurred during solar cycles 19--23, i.e., 1950-2009) was a rare or even unique event,
 in both magnitude and duration, in the past three millennia.
Except for these extreme cases, our reconstruction otherwise reveals that solar activity is
 well confined within a relatively narrow range.

\section{Modes of solar activity}

\begin{figure}
\centering \resizebox{7cm}{!}{\includegraphics{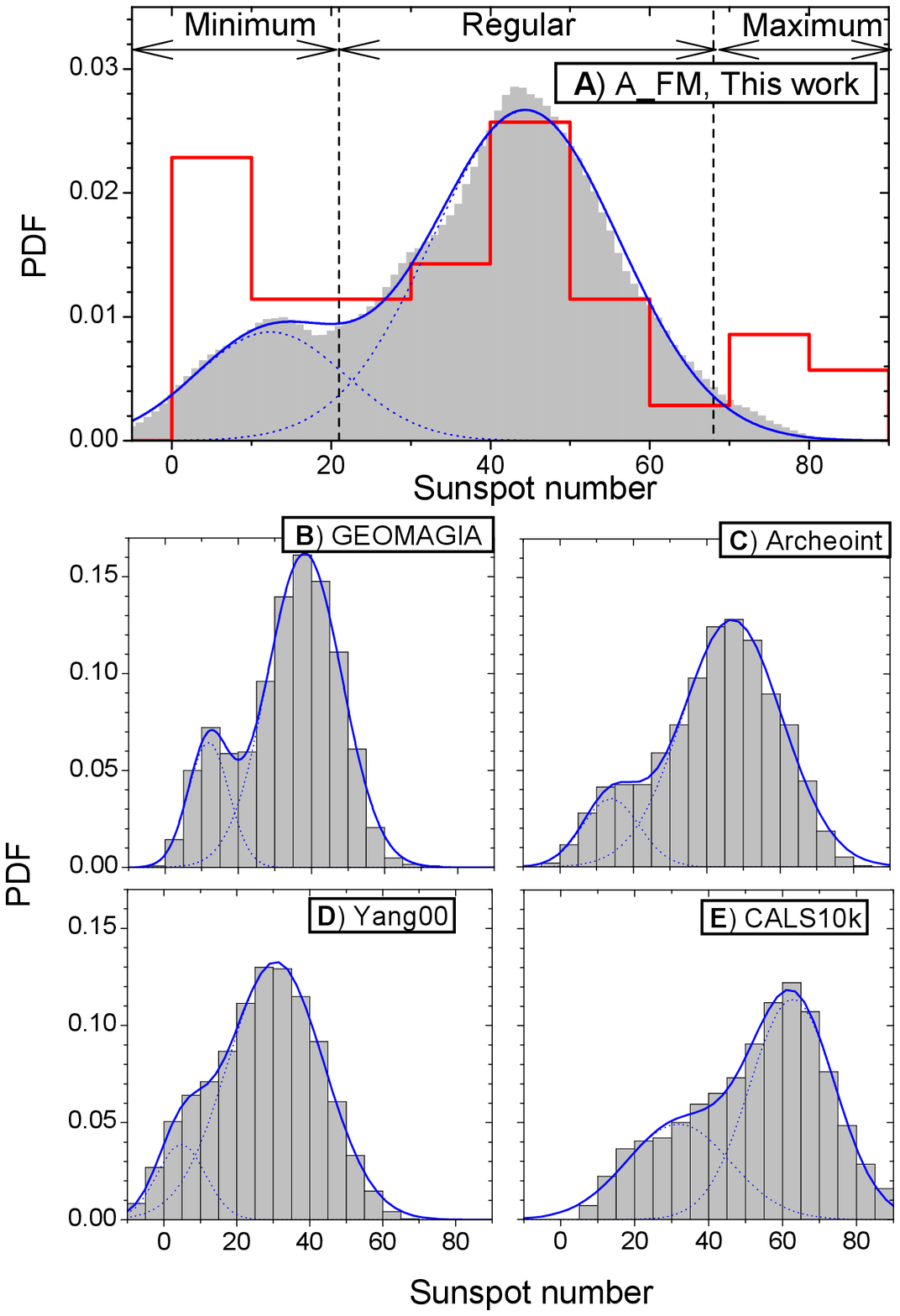}}
\caption{A) Probability density function (PDF) of the reconstructed decadal sunspot numbers as derived from the same $10^6$ series
 as in Fig.~\ref{Fig:Res} (gray-filled curve).
 The blue curve shows the best-fit bi-Gaussian curve (individual Gaussians with mean/$\sigma$ being 44/23 and 12.5/18 are shown as dashed blue curves).
 Also shown in red is the PDF of the historically observed decadal group sunspot numbers \citep{hoyt98} (using bins of width $\Delta$S=10).
 Vertical dashed lines indicate an approximate separation of the three modes and correspond to $\pm 1\sigma$ from the main peak, viz. S=21 and 67.
 Panels B and C depict PDFs built in the same way (but from a single mean reconstruction) using dipole moment models based on either the
  GEOMAGIA \citep{knudsen08} or ArcheoInt  \citep{genevey08} intensity databases.
 Panels D and E show PDFs (for the same time-period) inferred from a previous similar reconstruction \citep{usoskin_AA_07} using an early dipole
  moment model \citep{yang00} and from the recent reconstruction of \citet{roth13} using the CALS10k.1b model \citep{korte11}, respectively.
   }
\label{Fig:dist}
\end{figure}
Next we built a probability density function (PDF) of the reconstructed sunspot numbers (Fig.~\ref{Fig:dist}).
This PDF combines both the statistics intrinsically produced by the solar dynamo and all the errors considered above.
It clearly is bimodal, with a dip at sunspot number $S\approx 20$.
This reveals a main "regular" mode and a secondary "Grand minimum" mode.
The peak corresponding to the regular mode is in fact sharper than predicted by the corresponding best-fit Gaussian, which
 suggests a stronger
 separation between the Grand minimum and regular modes that is blurred by the statistics of reconstruction errors.
This distinct separation between modes is also suggested by the much longer time spent within each mode compared with both the correlation
 time of the signal and the time needed to switch from one mode to the other (Fig.~\ref{Fig:Res}).

The PDF built from the directly observed decadal group sunspot numbers \citep{hoyt98} (red line in Fig.~\ref{Fig:dist})
 has a central peak that coincides with the regular mode, consistent with the match seen in Fig.~\ref{Fig:Res} during
  the 1610--1950 AD overlapping period.
It also reveals two side peaks.
One is related to the 1645--1715 AD Maunder minimum, confirming a strong separation between Grand minimum and regular modes.
The other has no equivalent in the reconstructed sunspot numbers and is well above the $1\sigma$ regular mode threshold of $S=67$.
It suggests the possible existence of an additional "Grand maximum" mode, which the Sun experienced during solar cycles 19 through 23,
 before it shifted back to the regular mode during current solar cycle 24 (Fig.~\ref{Fig:Res}).
However, this PDF is based on small number statistics and does not provide compelling evidence for a
 Grand maximum mode separation \citep[cf.][]{usoskin_AA_07}.
A direct identification of such a Grand maximum mode in the reconstruction is indeed challenging.
According to the present result, no clear Grand maximum event occurred during the previous
 three millennia, although some episodes appear to have been very close to this (e.g., ca. 450 BC, 500 AD).
Previous reconstructions that cover longer time-periods \citep{solanki_Nat_04,steinhilber10,usoskin_AA_07} suggested that comparable
 events occurred earlier during the Holocene.
This suggests that Grand maxima, while not easily identified, are significantly less often experienced by the Sun than
 Grand minima (the latter occur about 16\% of the time, based on the present reconstruction).

\section{Assessment of systematic uncertainties}
\label{Sec:unc}

Here we assess model uncertainties.
Alternative reconstructions based on the ASD\_FM or ASDI\_FM ensembles of archeomagnetic field models
 \citep{licht13}, which were built using sediment paleomagnetic data in addition to the most robust archeomagnetic and volcanic data used
  in A\_FM, lead to identical conclusions (not shown).
Additionally, we performed sensitivity analyses with respect to systematic uncertainties whose influence is not accounted for by the
 ensemble of $10^{6}$ series in, namely key parameters of the carbon cycle model, the atmospheric cascade simulations, and
 the final conversion of $\phi$ into sunspot numbers.
Modifying these parameters may alter the overall variability amplitude of the reconstruction and/or introduce systematic offsets,
 but does not affect the identification of the distinct modes.
More generally, these tests confirm that no a posteriori adjustments are needed for the independently determined
 physical parameters used in the present reconstruction.

Most critical for identifying a distinct Grand minimum mode is the availability of a proper representation of
 the long-term trend in the geomagnetic field (Fig.~\ref{Fig:data}B), which affects the $Q$ values (Fig.~\ref{Fig:data}A),
 but is expected to be corrected for in the reconstructed $\phi(t)$ or $S(t)$ series (Fig.~\ref{Fig:Res}).
Indeed, previous reconstructions using earlier geomagnetic field models did not reveal any clear signature of distinct modes
 (see, e.g., Fig.~\ref{Fig:dist} panel D).
The difficulty of properly constraining long-term trends in geomagnetic field models may also be the reason why these previous
 reconstructions revealed hardly any signature of the Grand minimum mode, even when considering longer time periods \citep{usoskin_AA_07,steinhilber10,roth13}.
Reconstructions using geomagnetic models that heavily rely on sediment data, which are more prone to uncontrolled
 long-term biases (such as in Fig.~\ref{Fig:dist}E) than archeomagnetic and volcanic data, do not easily yield a distinct
 peak in the PDF, but a bump,
 corresponding to Grand minima, can still be observed with {\it a posteriori} knowledge.
In contrast, reconstructions using recent geomagnetic models based on the most reliable archeological and volcanic samples do lead
 to a clear mode separation (panels A--C of Fig-~\ref{Fig:dist}).

\section{Discussion and conclusions}

The cause of the strong variations in solar activity, and whether these result from pure random fluctuations, is still a subject of much debate
 \citep[e.g.][]{charbonenau10,choudhuri12,sokoloff04,moss08}.
While modern dynamo models can reproduce many observed features of the solar cycle, including secular variability and occurrence of
 seemingly random Grand minimum events \citep[e.g.][]{brandenburg08,moss08,choudhuri12}, this often requires a number of ad hoc
 assumptions \citep{pipin12}.
Thanks to the present reconstruction of solar activity, however, a much clearer picture of Grand minimum events now emerges.
These events cannot be described in terms of random fluctuations of a single solar-activity mode.
Instead, the evidence suggests that they are manifestations of a Grand minimum mode, distinct from the central regular mode, and that they
 occur as a result of sudden transitions from this central mode.
A similar pattern can be suggested for Grand maximum events as well, but the low statistics does not yet make this suggestion robust.
%{\bf
Transitions among these modes can then be understood either as transitions between different branches of the solar dynamo,
or as sudden changes in the governing parameters of this dynamo, as proposed by, for example, \citet{moss08}.
%}

The present result implies that

\noindent
$\bullet$ the Sun spends most of its time in a regular activity mode with a range of cycle-averaged sunspot numbers of
 between about 20 and 67

\noindent
$\bullet$
the Grand minimum state clearly is a separate mode of activity, within which the Sun spends about 1/6 of its time

\noindent
$\bullet$
there is an indication that the Grand maximum also corresponds to a separate mode of activity, but the low statistics does
 not allow us to firmly conclude on this, yet.

\vskip 0.2cm
\noindent
These observations provide important constraints for solar-activity models, and will help to deepen
 our understanding of the processes that drive solar and stellar variability.

\begin{acknowledgements}
The data reported in the paper are presented in the online material.
This work was initiated and partly supported by IPGP visiting Program.
I.U. also acknowledges support from the University of Oulu.
G.H., Y.G., A.L., E. T. and A. K. were supported by IPGP and CNRS.
R.R. and F.J. acknowledge support by the Swiss National Science Foundation and the EU project Past4Future (Grant No 243908).
G.K. was partly supported by the Program No. 22 presidium RAS and by the Academy of Finland.
Support by the Academy of Finland to the ReSoLVE Center of Excellence (project no. 272157) is acknowledged.
This is IPGP contribution No. 3487. 
\end{acknowledgements}

%\bibliographystyle{aa} % style aa.bst
%\bibliography{J:/usoskin/papers/usoskin_all}
%\bibliography{C:/DATA_/USOSKIN/papers/usoskin_all}
%\bibliographystyle{plainnat}

\end{document}